\documentclass[reprint,amsmath,amssymb, aps]{revtex4-2}

\usepackage{gensymb}
\usepackage{graphicx}
\usepackage{dcolumn}
\usepackage{bm}
\usepackage{subcaption}
\usepackage{subfloat}
\usepackage{color}
\usepackage{bm}
\usepackage{ulem}
\usepackage{mhchem} 
\usepackage{tikz}
\usetikzlibrary{plotmarks}
\newcommand\marksymbol[2]{\tikz[#2,scale=1.2]\pgfuseplotmark{#1};}

\definecolor{brown(web)}{rgb}{0.65, 0.16, 0.16}
\definecolor{blue_green}{rgb}{0.0, 0.75, 0.75}

\def\red{\textcolor{red}}

\def\blue{\textcolor{blue}}
\def\violet{\textcolor{violet}}
\def\purple{\textcolor{purple}}
\def\orange{\textcolor{orange}}

\begin{document}

\preprint{APS/123-QED}

\title{Controlling Viscous Fingering Instabilities of Complex Fluids}

\author{Alban Pouplard}
\author{Peichun Amy Tsai$^\ast$}%
\affiliation{%
 Department of Mechanical Engineering, University of Alberta,\\ Edmonton, Alberta, Canada T6G 2G8
}%

\date{\today}


\maketitle


{\bf The process of one fluid pushing another is universally common while involving complex interfacial instabilities. Particularly, occurring in a myriad of natural and industrial processes, 
wavy fingering patterns frequently emerge when a less viscous fluid pushes another more viscous one, such as water invading oil, in a porous medium. Such finger-shaped interfaces producing partial displacement significantly affect the efficiency of numerous applications, for example, chromatography \cite{Fernandez1996}, printing devices \cite{Pitts1961}, coating flows~\cite{Grillet1999}, oil-well cementing, as well as large-scale technologies of groundwater and enhanced oil recovery (EOR) \cite{Green2018}. This classical viscous fingering instability \cite{Paterson1981, Paterson1985, Saffman1986, Homsy1987} is notoriously difficult to control because the two fluids' viscosity or mobility ratio is often fixed and yet the predominant drive of the instability. Although some strategies have been recently revealed for simple fluids of constant viscosity, the feasibility of controlling the fundamental viscous fingering instability for omnipresent complex fluids has not been established. Here, we demonstrate how to control a common complex fluid (of a power-law fluid with a yield-stress) using a narrow tapered cell theoretically and experimentally.}

The unfavorable mobility or viscosity contrast commonly triggers fingering patterns during immiscible fluid-fluid displacement in a porous media, hindering a full swipe of the displaced fluid. This so-called viscous fingering (VF) or Saffman-Taylor instability \cite{Paterson1981, *Paterson1985, Saffman1986, Homsy1987, *Chen1987} has been extensively studied since the 1980s, particularly with a convenient paradigm of Hele$-$Shaw cells consisting of two parallel plates spaced with a constant gap thickness. 
Recent studies using simple fluids have considered centrifugally driven VF via rotation \cite{Dias2013_2} and found that the inertia tends to increase the finger-width \cite{Chevalier2006} and curvature-dependent surface tension can theoretically lead to the stabilization (destabilization) of conventionally unstable (stable) situations \cite{Rocha2013}.  In the last two decades, studies of viscous fingering have been extended to complex fluids, usually leading to wider fingers compared to the simple Newtonian counterparts \cite{Park1994}. 
Besides, intriguing side-branching patterns with multiple small sided-fingers are often observed with complex, yield-stress fluids \cite{Coussot1999}.

The control of the fingering instabilities plays a significant role in enhancing the efficiency of various industrial applications. For simple Newtonian fluids, several strategies have recently been developed to suppress the fingering instability, for example, using time-dependant injection flow rate \cite{Cardoso1995, Dias2012, Zheng2015}, an elastic confinement \cite{Pihler2012, *Pihler2013, Housseiny2013}, a gap-gradient cell \cite{Housseiny2012, Stone2013, Bongrand2018}, and an external electric field \cite{Mirzadeh2017}. Nevertheless, such control of the primary VF instability has not been reported for complex fluids, which are commonly present in natural and industrial settings.  Here we demonstrate the feasibility of inhibiting the viscous fingering instability of complex, yield stress fluids using a radially-tapered narrow cell by carrying out experiments and linear stability analysis.

\begin{figure*} [htb!]
    \begin{center}
    \includegraphics[width=6.5in]{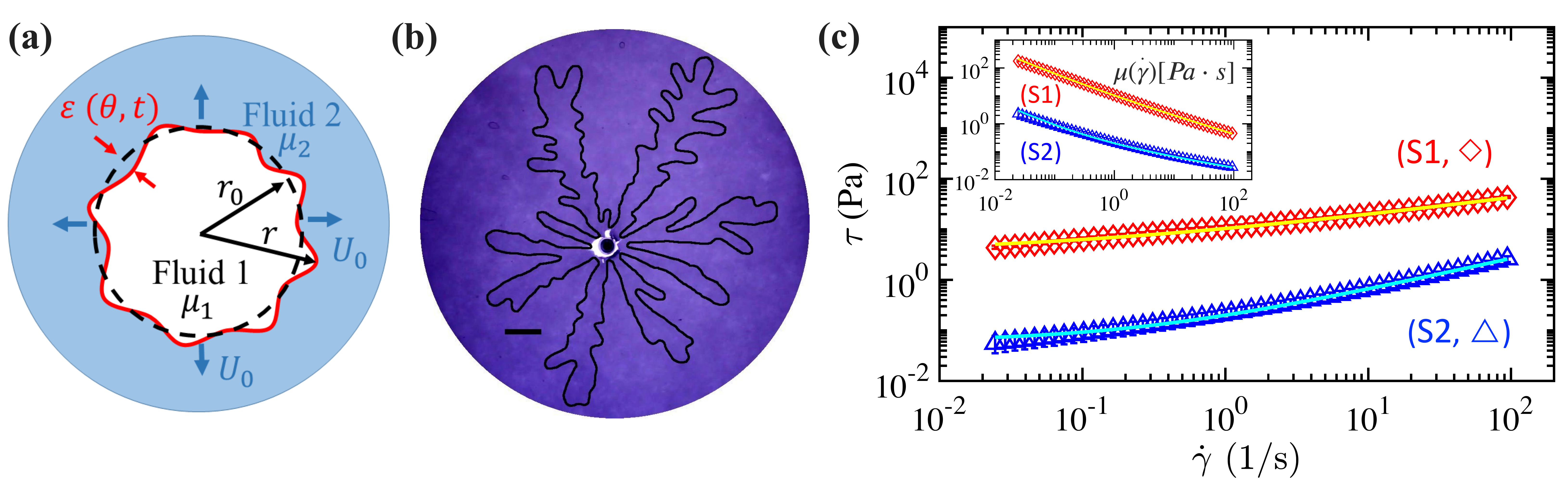}
    \end{center}
    \protect{\caption{ (a) Top-view schematics of the fluid-fluid displacement experiment where one less viscous complex fluid 1 of varying viscosity ($\mu_1$) with shear rate ($\dot{\gamma}$) is pushing another immiscible one, denoted as complex fluid 2 with changing viscosity ($\mu_2$). (b) Representative experimental snapshot of complex viscous fingering produced with a complex yield-stress (PAA) solution (S1) displaced by a gas injected with a flow rate $Q = 0.2$ slpm in a flat Hele-Shaw cell, under the mobility contrast ${\cal M} = \mu_2/\mu_1 = 5.58 \times 10^4$ and at the interface velocity $U_0 = 14.3$ mm/s. The scale bar corresponds to 2~cm. (c) Flow curves of shear stress, $\tau$, and viscosity, $\mu$, depending on $\dot{\gamma}$ for the two complex solutions used: (S1, \red{$\Diamond$}) and (S2, \blue{$\bigtriangleup$}). The lines in (c) correspond to the best fits of the data to the Herschel-Bulkley model [Eq.~\eqref{eq1_HB}].
    }
    \label{Fig1}}
\end{figure*}

Experimentally, we use two different aqueous solutions of PolyAcrylic Acid solution (PAA) as a wetting yield-stress fluid (see Methods). We first fill in one complex PAA solution in a radial cell and subsequently inject a gas (nitrogen, viscosity $\mu_1 = 1.76 \times 10^{-5}$ Pa$\cdot$s at $20~\degree$C) as a pushing fluid 1 (see Fig.~\ref{Fig1}a-b). The gas is injected at a constant flow rate, $Q$, ranging from $0.02$ to $2$~slpm (standard liter per minute) by a flow controller (Alicat). 
Fig.~\ref{Fig1}(c) shows the rheological measurements (AntonPaar MCR302) of the shear stress ($\tau$) varying with shear rate ($\dot{\gamma}$) for the two complex solutions. Neglecting the elastic properties (see Methods for the justification), the flow curve data shows an excellent fit with the common Herschel$-$Bulkley (HB) model \cite{Herschel1926}:
\begin{align} \label{eq1_HB}
\tau  = \tau_{c} + \kappa \dot{\gamma}^{n},
\end{align}
where $\tau_{c}$, $\kappa$ and $n$ correspond to the yield stress, the consistensy index, and the power-law index, respectively. 
Shown in Fig.~\ref{Fig1}(c) inset, the viscosity data ($\mu$) varying with $\dot{\gamma}$ is well described by the corresponding HB model [Eq.~\eqref{eq1_HB}] via $\mu \equiv \tau/\dot{\gamma} = \tau_{c}/\dot{\gamma} + \kappa {\dot{\gamma}}^{n-1}$. Table~\ref{table1} summarizes the best nonlinear-fit results of the rheological measurements of $\tau = f(\dot{\gamma})$ for $0.025 \leq \dot{\gamma} \leq$ 86~s$^{-1}$. The corresponding HB fitting functions are plotted as lines in Fig.~\ref{Fig1}(c). 
Both solutions are shear-thinning, with decreasing viscosity with increasing $\dot{\gamma}$, i.e., $n < 1$. However, the neutralized PAA solution with \ce{NaOH} (S1) is more viscous, by $\approx 15\times - 70\times$ than (S2) without \ce{NaOH} depending on $\dot{\gamma}$, and has a greater $\tau_c$ but a smaller $n$.

\begin{table}[htb!]
\caption{Rheological parameters for the two complex, yield-stress fluids with the HB model [Eq.~\eqref{eq1_HB}].
}
\label{table1}
\begin{ruledtabular}
\begin{tabular}{l c c c c c}
\textrm{Yield-stress} & 
\textrm{PAA} & 
\textrm{NaOH}&
\textrm{$\tau_{c}$}&
\textrm{$\kappa$}&
\textrm{$n$}\\
\textrm{solution} &
\textrm{(wt~\%)} &
\textrm{(wt~\%)} &
\textrm{(Pa)} &
\textrm{(Pa$\cdot$s$^n$)} \\
\colrule
(S1) & 0.10 & 0.034 & 3.2857 & 7.1179 & 0.3721 \\
(S2) & 0.10 & 0 & 0.0596 & 0.1413 & 0.6333 \\
\end{tabular}
\end{ruledtabular}
\end{table}


Using flat Hele$-$Shaw cells, we observe complex fingering patterns, which overall resemble the classical viscous fingering for simple Newtonian fluids but has complex side-fingers along the side of the major fingers (shown in Fig.~\ref{Fig1}b). In agreement, similar patterns were observed previously, referred as side-branching \cite{Jirsaraei2005} or the elasto-inertial regime \cite{Eslami2017, Eslami2019}. Interestingly, the side-branched fingers can be obtained only at high flow-rates ($Q \geq 1.5$ slpm) for the fluid (S2) but for all the experimental range of $Q = 0.02-1.5$ slpm for the more-viscous (S1).

 \begin{figure*}[ht!]
    \begin{center}
    \includegraphics[width=7in]{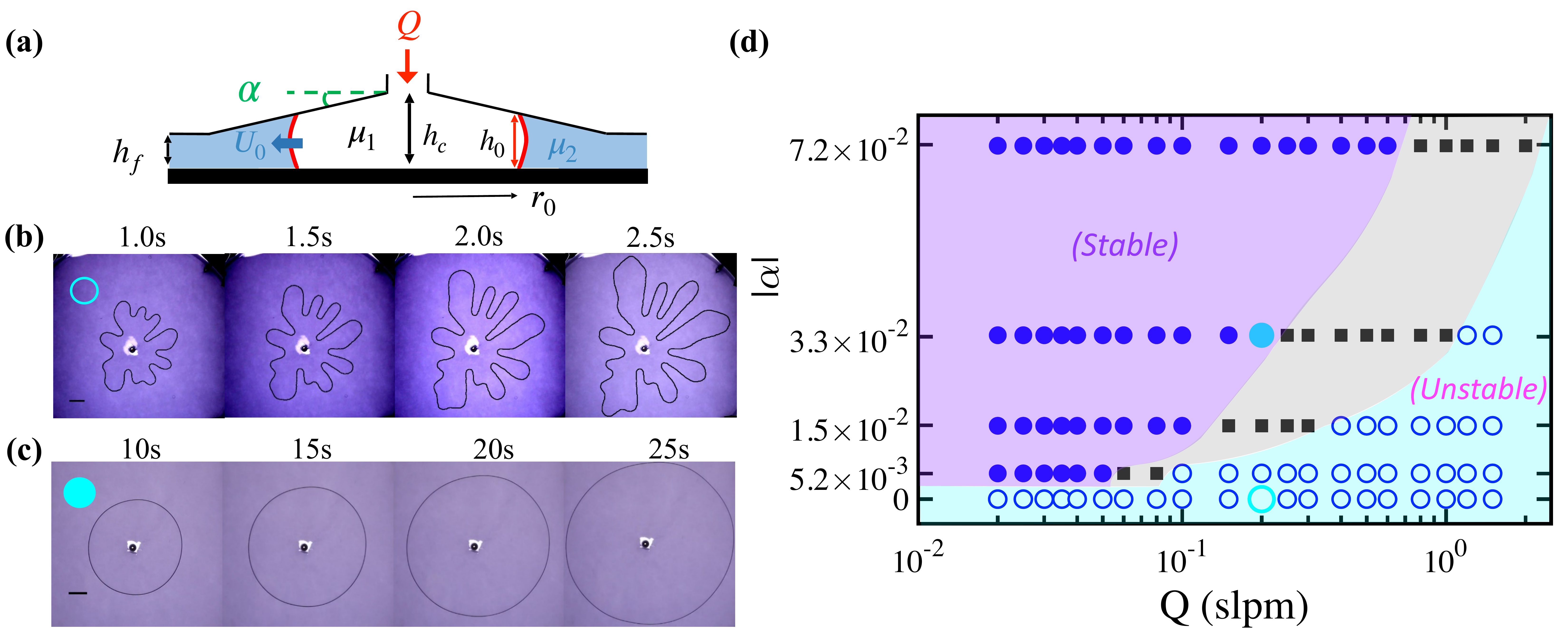}
    \end{center}
    \protect{\protect{\caption{{\bf Control of complex viscous fingering} using a radially-tapered cell, with a linearly varying gap thickness ($h = h_c + \alpha r$), schematically shown in (a), the side-view of the experiment.
    (b) Experimental snapshots of a branched viscous fingering pattern observed when a gas is pushing the complex solution (S2) in a flat Hele-Shaw cell with $h_c = 0.5$~mm and $Q = 0.2$~slpm. (c) By contrast, snapshots of a stable interface obtained when the gas is pushing (S2) in a tapered cell of the gap gradient $\alpha = -3.33 \times 10^{-2}$, with $h_c = 5.16$~mm and $Q = 0.2$~slpm. The scale bars in (b) and (c) represent $20$~mm.(d) Experimental results of stability diagram, with uniform stable (filled circle) vs. fingering/wavy unstable interfaces (open circle) under various values of flow rate, $Q$, and the tapered gap gradient, $|\alpha|$. Black squares ($\blacksquare$) represent a transitional state where the interface starts to develop a wavy profile.
    \label{Fig2}}}}
\end{figure*}

When using converging cells, we stopped observing side-branched fingers but smooth classical viscous fingers with Q = $0.02-1.5$ slpm. This is consistent with a recent experimental study \cite{Eslami2020_1}, revealing mitigation of side-branching (but not inhibition of the primary VF) for a complex yield-stress fluid in a rectangular tapered cell. 
Remarkably, with suitable rheological and flow parameters, we can control and inhibit the primary fingering instability and observe complete stable and flat interfaces between the pushing Newtonian gas and the yield-stress fluid, as illustrated by Fig.~\ref{Fig2}c.

The first crucial observation is that the fluid-fluid interface is stable at a lower $Q$ when keeping $\alpha$ and $h_c$ constant. In Fig.~\ref{Fig2}(d) phase diagram under various values of $\alpha$ and $Q$, we differentiate three types of displacements observed during the experiments with the fluid (S2), namely uniform stable (filled circle, $\marksymbol{*}{blue}$), fingering/wavy unstable (open circle, $\marksymbol{o}{blue}$), and transitional (filled square, $\marksymbol{square*}{black}$) displacements. The latter corresponds to the transitional state where the interface starts to develop a wavy profile.
A converging gap gradient helps the interface stabilize, and the transition from stable to unstable interfaces happens at a higher flow rate. The stability diagram is established only for the complex fluid (S2) since a complete sweep has never been observed with the more-viscous fluid (S1) of a high mobility contrast, ${\cal M} = \mu_2/\mu_1$ = $2.68\times 10^{4} - 1.16\times 10^{7}$. The complex fluid (S2) has a lower and small ${\cal M}$-range of $1.61\times 10^{3} - 1.67\times 10^{5}$ than (S1). These contrast results between the two complex solutions highlight not only the complexity but also the importance of rheological parameters, via $\kappa$, $n$ and local $\dot{\gamma}$, in controlling complex viscous fingering.


To gain physical insights, we develop a linear stability analysis generalized to two yield-stress, power-law fluids (Fluid 1 pushing Fluid 2) in a radially-tapered cell, as depicted in Fig.~\ref{Fig2}a.  The introduction of a constant gap gradient ($\alpha$) produces a linearly-varying height between the two plates of the cell. Considering the fluids' interface at $r = r_0$, the height $h$ can be expressed as $h(r) = h_0 + \alpha \left(r-r_0 \right)$, with $h_0$ the gap-thickness at the interface. For fluids in the narrow gap, based on the lubrication theory, we use the effective Darcy's law replacing the constant viscosity, $\mu$, by the effective shear-dependent viscosity, $\mu_\text{eff}$. Neglecting the fluids' elastic properties~\cite{Coussot1999}, the governing equations of the immiscible, complex fluids are the continuity equation (taking gap-variation into account) and 2D depth-average Darcy's law:
\begin{equation} \label{eq.2.appxB}
\nabla \cdot \left(h {\bf U}_j \right) = 0 \quad \text{and} \quad
{\bf U}_j = -\frac{h^2}{12\mu_{\text{eff}j}} \vec{\nabla} P_j,
\end{equation}
where ${\bf U}_j(r,\theta) = (u_{rj} , u_{\theta j})$ and $P_j(r,\theta)$ are the depth-average velocity and pressure fields of the fluid indexed $j$, respectively. $j$ represents the two complex fluids during the fluid-fluid displacement process; $j = 1$ (2) denotes the pushing (displaced) fluid.

The complex fluid's viscosity ($\mu_{\text{eff} j}$) is modeled using the Herschel-Bulkley law [see Eq.~\eqref{eq1_HB}] for yield-stress fluids, with the local shear rate $\dot{\gamma} = u_{rj}/h$, and expressed as:
$\mu_{\text{eff} j} = \frac{\tau_{cj}}{\overset{.}{\gamma}} + \kappa_j {\overset{.}{\gamma}}^{n_j-1}$, with yield stress $\tau_{cj}$, consistensy index ($\kappa_j$), and power-law index ($n_j$). The depth-average continuity Eq. can be expressed using the pressure field ($P_j$) and further simplified.  By setting $n_j = 1$ and $\tau_{cj} = 0$, we obtain and recover the simple Newtonian fluid case: $\frac{\partial^2 P_j}{\partial r^2} + \frac{1}{r} \frac{\partial P_j}{\partial r} + \frac{3 \alpha}{h} \frac{\partial P_j}{\partial r} + \frac{1}{r^2} \frac{\partial^2 P_j}{\partial \theta^2} = 0$, reported by Al-Housseiny and Stone \cite{Stone2013}.


In the linear stability analysis, the pressure field is expressed as the solutions of the base state and the perturbation, $\epsilon (\theta,t) = \epsilon_0 r_0(t) \exp{ \left(i k \theta+ \sigma t \right)}$:
\begin{align} 
P_j (r,\theta,t) = f_j (r) + g_{kj} (r) \epsilon (\theta,t), \label{eq.8.appxB}
\end{align}
where $f_j(r)$ corresponds to the base-state pressure when the interface is stable and independent of $\theta$. The term of $g_{kj}(r) \epsilon$ represents the perturbation that propagates along the interface with wavenumber ($k$) and the growth rate of the perturbation ($\sigma$). We employ the kinematic boundary conditions, i.e., two complex fluids moving at the same velocity at the interface, and the Young$-$Laplace equation for the pressure jump at the interface due to surface tension and curvature. 
To obtain analytical solutions, we assume that the fluid yield stress is negligible compared to the viscous stress, i.e., small Bingham ($Bn_j \ll 1$) situation, where the Bingham number is the ratio of the yield to viscous stress: $Bn_j= \frac{\tau_{cj}}{\kappa_j \left(\frac{u_{rj}}{h} \right)^{n_j}}$.
Focusing on the moment when the perturbation starts to propagate, implying small perturbation $\left(\epsilon \ll 1 \right)$, $\epsilon_0 \ll 1$, $g'_{kj}(r) \epsilon \ll f'_j(r)$, and negligible high-order terms of $O \left(\epsilon^2 \right)$, we obtain the dimensionless dispersion relation with the dimensionless growth rate, $\bar{\sigma} = \frac{\sigma r_0}{U_0}$, and the dimensionless wavenumber,
$\bar{k} = k$  (see Supplementary Information):
\begin{widetext}
\begin{equation} \label{eq.24.appxB}
\begin{split}
 \frac{12 \overline{\sigma} h_0}{\gamma} \left(\kappa_1 \sqrt{n_1} \left(\frac{U_0}{h_0} \right)^{n_1} + \kappa_2 \sqrt{n_2} \left(\frac{U_0}{h_0} \right)^{n_2} \right) &= -\frac{12 U_0}{\gamma}
 \left(\sqrt{n_1} \mu_1|_{r=r_0} + \sqrt{n_2} \mu_2|_{r=r_0} \right)  - \frac{12 \alpha r_0}{\gamma} 
\biggl( 2 \sqrt{n_1} \tau_{c1} + 2 \sqrt{n_2} \tau_{c2} \biggl)\\
 - \frac{12 \alpha r_0}{\gamma} \biggl( \sqrt{n_1} \kappa_1 \left( \frac{U_0}{h_0} \right)^{n_1} + \sqrt{n_2} \kappa_2 \left( \frac{U_0}{h_0} \right)^{n_2} \biggl)
& + \overline{k} \left(\frac{12 U_0}{\gamma} \left( \mu_2|_{r=r_0} - \mu_1|_{r=r_0} \right) + 2 \alpha \cos{\theta_c} + \frac{{h_0}^2}{{r_0}^2} \right) - \frac{{h_0}^2}{{r_0}^2} {\overline{k}}^3.
\end{split}
\end{equation}
\end{widetext}

\begin{figure*}[htb!]
    \begin{center}
    \includegraphics[width=5.5in]{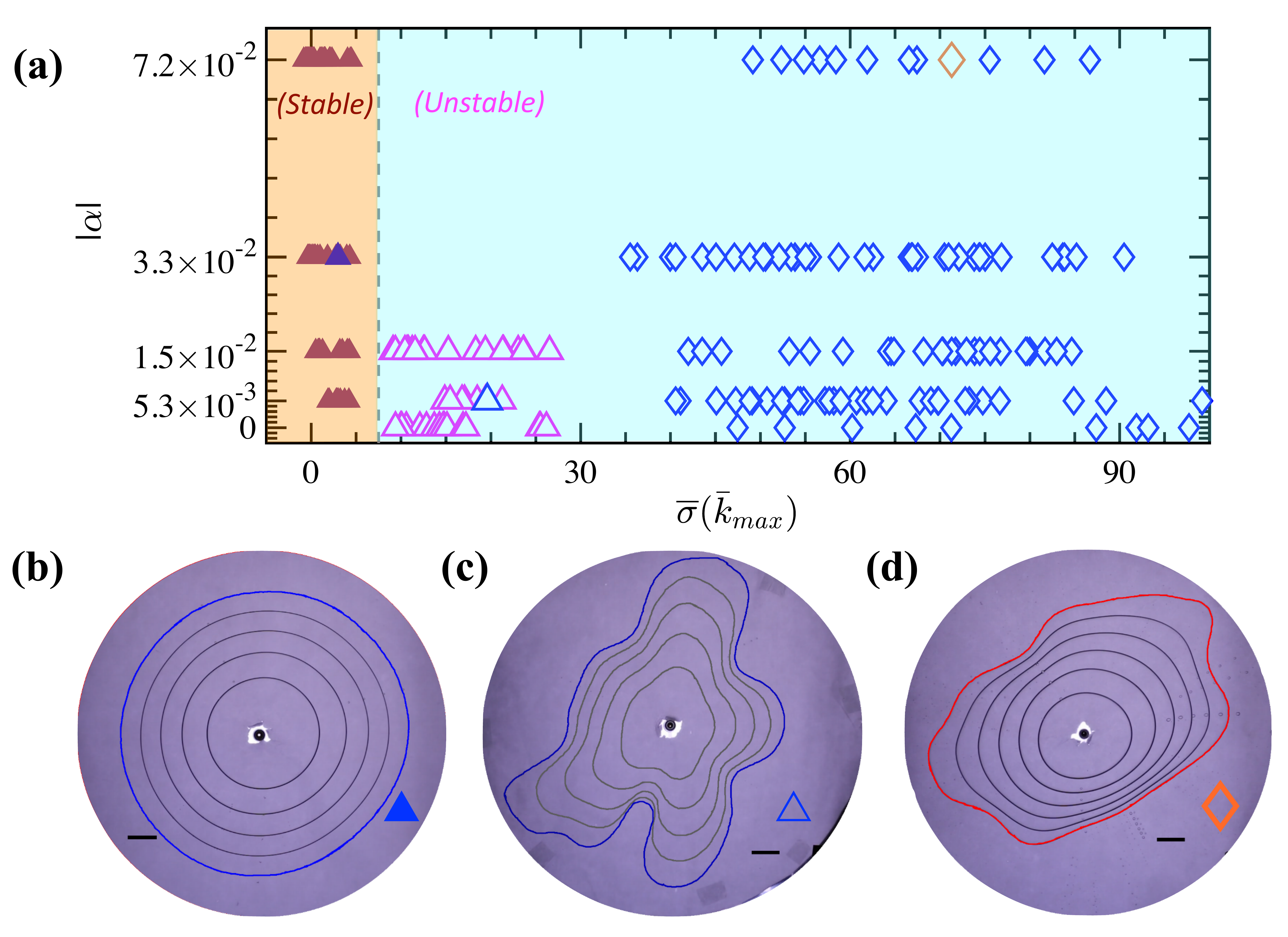}
    \end{center}
    \caption{{\bf Comparison between experimental and theoretical results}: (a) The growth rate of the perturbation at the most unstable mode of $k_{max}$, $\overline{\sigma}(\bar{k}_{max})$ using Eq.~\eqref{eq.24.appxB} and \eqref{eq.25.appxB} for different experiments performed with various gap-gradient, $|\alpha|$.
   The experiments with the more viscous solution (S1) always show unstable wavy interface ({\large\blue{$\Diamond$}}). In contrast, for the less viscous complex fluid (S2), stable displacement  ({{\large\purple{$\blacktriangle$}}) and unstable interface}({{\large\violet{$\vartriangle$}}) are observed with nearly-zero and relatively-large growth rate $\overline{\sigma}$, respectively. (b)-(d) are the overlays of experimental snapshots, revealing the evolution of the interface profiles for the three big symbols ({\large\blue{$\blacktriangle$}}, {\large\blue{$\vartriangle$}}, {\large\orange{$\Diamond$}}) in (a), respectively.  
   The time steps are $\delta t=22$~s, 0.6~s, and 1~s in (b), (c) and (d), respectively. Each scale-bar represents a length scale of 20~mm. \label{Fig3}}}
\end{figure*}

Consequently, taking $\tau_{cj} =0$, $n_j = 1$, $\kappa_j = \mu_j$ for simple Newtonian fluids, and defining $\lambda = \frac{\mu_1}{\mu_2}$ and $Ca=\frac{12 U_0 \mu_2}{\gamma}$, the dispersion relation recovers to the same formula by Al-Housseiny and Stone for Newtonian fluids with constant viscosity [Eq. (24) in \cite{Stone2013}]. In addition to the crucial influences of $\alpha$, $\lambda$, $Ca$, and wetting angle ($\theta_c$) for the simple fluid case, the derived dimensionless perturbation growth rate ($\overline{\sigma}$) as a function of ($\overline{k}$) depends on the fluids' rheological properties ($\kappa_j, n_{j}, \tau_{cj}$) and the local velocity, radius, and gap thickness at the interface ($U_0, r_0, h_0$, respectively) for the complex yield-stress fluids.  Besides, if the pertubation's growth rate is less than zero for every wavenumber, $k$, the interface will always be stable theoretically. The wavenumber at the maximum growth ($\bar{k}_{max}$) can be found by taking the derivative of the above dimensionless dispersion Eq.~\eqref{eq.24.appxB} w.r.t. $\bar{k}$ and setting $\frac{ \partial \overline{\sigma}}{\partial \overline{k}} = 0$:
\begin{equation} \label{eq.25.appxB}
\bar{k}_{max} = \left(\frac{\frac{{h_0}^2}{{r_0}^2} + 2 \alpha \cos{\theta_c} + \frac{12 U_0}{\gamma} \left(\mu_2|_{r=r_0} - \mu_1|_{r=r_0} \right)}{3 \frac{{h_0}^2}{{r_0}^2}} \right)^{\frac{1}{2}}.
\end{equation}

Using the wavenumber of maximum growth $\bar{k}_{max}$ (Eq.~\eqref{eq.25.appxB}) and 
(Eq.~\eqref{eq.24.appxB}), we obtain the growth rate at the most unstable mode. To compare with our theoretical prediction, taking the values of viscosity ($\mu_1$ and $\mu_2$), $U_0$, $r_0$ and $h_0$ from the experiments, we plotted the $\overline{\sigma}(\bar{k}_{max})$ from Eq.~\eqref{eq.24.appxB} in Fig.~\ref{Fig3}a. From the graph, we can observe a transition from stable to unstable interfaces when $\overline{\sigma}(\bar{k}_{max}) < 7.5$ from the experimental data, slightly deviating from the theoretical value of $\overline{\sigma} < 0$. The deviation between our experimental results and the theory may be due to the few assumptions we made. The impact of the gravity and the elastic properties have been ignored. Moreover, whenever $|\alpha|$ is getting bigger, the assumptions of small ratio of gap change ($ \frac{\alpha (r-r_0)}{h_0} \ll 1$) as well as the characteristic length scale over which the depths varies being much larger than that of the perturbation scale ($\frac{k h_0}{\alpha r_0} \gg 1$) might be unjustified. Last but not the least, we neglected the yield stress compared to the viscous stress by assuming small $Bn \ll 1$. These assumptions likely contribute to the drifting of the critical maximum growth rate from $0$ (theoretically) to $7.5$.

In summary, we have demonstrated a powerful way of stabilizing the primary viscous fingering instability for complex yield-stress fluids using a tapered narrow cell for the first time. Experimentally, using a radially-tapered cell, we can hinder complex fingering patterns, e.g., eliminating side-fingers for the more-viscous (S1) and suppressing wavy interfaces completely for the less-viscous complex fluid (S2). With a linear stability analysis using the effective Darcy's law, we derive the dispersion relation and establish a convenient stability criterion corresponding to the perturbation's growth rate of the most unstable mode. In addition to the viscosity contrast ($\lambda$), gap gradient ($\alpha$), $\theta_c$ and Capillary number ($Ca$) for the simple Newtonian fluids, several vital parameters affect the complex fluids' viscous fingering stability criterion, namely the fluid's rheological characteristics, such as $\kappa$, $\tau_c$, $n$, and $\gamma$, as well as the interface position, gap thickness, and velocity ($r_0$, $h_0$ and $U_0$, respectively). This theoretical stability criterion through $\overline{\sigma}(\bar{k}_{max})$, despite the assumption of small $Bn \ll 1$, shows fair agreement with the experimental results with two yield-stress fluids of distinct mobility ratios. These results, particularly the complex dispersion relation [Eq.~\eqref{eq.24.appxB}] and $\overline{\sigma}(\bar{k}_{max})$, provide quantitative insights into the designs and strategies for controlling viscous fingering and interfacial profiles during complex fluids' displacement in microfluidics, narrow cells, packed beads, and porous media.

\begin{acknowledgments}
{A.P.} and {P.A.T.} thankfully acknowledge the funding support from the Natural Sciences and Engineering Research Council of Canada (NSERC) Discovery grant (RGPIN-2020-05511). {P.A.T.} holds a Canada Research Chair in Fluids and Interfaces (CRC TIER2 233147). This research was undertaken, in part, thanks to funding from the Canada Research Chairs (CRC) Program.
\end{acknowledgments}

$^\ast$ Email address of the corresponding author: 
\\P. A. Tsai (peichun.amy.tsai@ualberta.ca).


\bibliography{Ref}

\providecommand{\noopsort}[1]{}\providecommand{\singleletter}[1]{#1}%
\begin{thebibliography}{29}%
\makeatletter
\providecommand \@ifxundefined [1]{%
 \@ifx{#1\undefined}
}%
\providecommand \@ifnum [1]{%
 \ifnum #1\expandafter \@firstoftwo
 \else \expandafter \@secondoftwo
 \fi
}%
\providecommand \@ifx [1]{%
 \ifx #1\expandafter \@firstoftwo
 \else \expandafter \@secondoftwo
 \fi
}%
\providecommand \natexlab [1]{#1}%
\providecommand \enquote  [1]{``#1''}%
\providecommand \bibnamefont  [1]{#1}%
\providecommand \bibfnamefont [1]{#1}%
\providecommand \citenamefont [1]{#1}%
\providecommand \href@noop [0]{\@secondoftwo}%
\providecommand \href [0]{\begingroup \@sanitize@url \@href}%
\providecommand \@href[1]{\@@startlink{#1}\@@href}%
\providecommand \@@href[1]{\endgroup#1\@@endlink}%
\providecommand \@sanitize@url [0]{\catcode `\\12\catcode `\$12\catcode
  `\&12\catcode `\#12\catcode `\^12\catcode `\_12\catcode `\%12\relax}%
\providecommand \@@startlink[1]{}%
\providecommand \@@endlink[0]{}%
\providecommand \url  [0]{\begingroup\@sanitize@url \@url }%
\providecommand \@url [1]{\endgroup\@href {#1}{\urlprefix }}%
\providecommand \urlprefix  [0]{URL }%
\providecommand \Eprint [0]{\href }%
\providecommand \doibase [0]{https://doi.org/}%
\providecommand \selectlanguage [0]{\@gobble}%
\providecommand \bibinfo  [0]{\@secondoftwo}%
\providecommand \bibfield  [0]{\@secondoftwo}%
\providecommand \translation [1]{[#1]}%
\providecommand \BibitemOpen [0]{}%
\providecommand \bibitemStop [0]{}%
\providecommand \bibitemNoStop [0]{.\EOS\space}%
\providecommand \EOS [0]{\spacefactor3000\relax}%
\providecommand \BibitemShut  [1]{\csname bibitem#1\endcsname}%
\let\auto@bib@innerbib\@empty
\bibitem [{\citenamefont {Fernandez}\ \emph {et~al.}(1996)\citenamefont
  {Fernandez}, \citenamefont {Norton}, \citenamefont {Jung},\ and\
  \citenamefont {Tsavalas}}]{Fernandez1996}%
  \BibitemOpen
  \bibfield  {author} {\bibinfo {author} {\bibfnamefont {E.~J.}\ \bibnamefont
  {Fernandez}}, \bibinfo {author} {\bibfnamefont {T.~T.}\ \bibnamefont
  {Norton}}, \bibinfo {author} {\bibfnamefont {W.~C.}\ \bibnamefont {Jung}},\
  and\ \bibinfo {author} {\bibfnamefont {J.~G.}\ \bibnamefont {Tsavalas}},\
  }\bibfield  {title} {\bibinfo {title} {A column design for reducing viscous
  fingering in size exclusion chromatography},\ }\href@noop {} {\bibfield
  {journal} {\bibinfo  {journal} {Biotechnol. Prog.}\ }\textbf {\bibinfo
  {volume} {12}},\ \bibinfo {pages} {480} (\bibinfo {year} {1996})}\BibitemShut
  {NoStop}%
\bibitem [{\citenamefont {Pitts}\ and\ \citenamefont
  {Greiller}(1961)}]{Pitts1961}%
  \BibitemOpen
  \bibfield  {author} {\bibinfo {author} {\bibfnamefont {E.}~\bibnamefont
  {Pitts}}\ and\ \bibinfo {author} {\bibfnamefont {J.}~\bibnamefont
  {Greiller}},\ }\bibfield  {title} {\bibinfo {title} {The flow of thin liquid
  films between rollers},\ }\href@noop {} {\bibfield  {journal} {\bibinfo
  {journal} {J. Fluid Mech.}\ }\textbf {\bibinfo {volume} {11}},\ \bibinfo
  {pages} {33} (\bibinfo {year} {1961})}\BibitemShut {NoStop}%
\bibitem [{\citenamefont {Grillet}\ \emph {et~al.}(1999)\citenamefont
  {Grillet}, \citenamefont {Lee},\ and\ \citenamefont {Shaqfeh}}]{Grillet1999}%
  \BibitemOpen
  \bibfield  {author} {\bibinfo {author} {\bibfnamefont {A.~M.}\ \bibnamefont
  {Grillet}}, \bibinfo {author} {\bibfnamefont {A.~G.}\ \bibnamefont {Lee}},\
  and\ \bibinfo {author} {\bibfnamefont {E.~S.}\ \bibnamefont {Shaqfeh}},\
  }\bibfield  {title} {\bibinfo {title} {Observations of ribbing instabilities
  in elastic fluid flows with gravity stabilization},\ }\href@noop {}
  {\bibfield  {journal} {\bibinfo  {journal} {J. Fluid Mech.}\ }\textbf
  {\bibinfo {volume} {399}},\ \bibinfo {pages} {49} (\bibinfo {year}
  {1999})}\BibitemShut {NoStop}%
\bibitem [{\citenamefont {Green}\ and\ \citenamefont
  {Willhite}(2018)}]{Green2018}%
  \BibitemOpen
  \bibfield  {author} {\bibinfo {author} {\bibfnamefont {D.~W.}\ \bibnamefont
  {Green}}\ and\ \bibinfo {author} {\bibfnamefont {G.~P.}\ \bibnamefont
  {Willhite}},\ }\href@noop {} {\emph {\bibinfo {title} {Enhanced oil
  recovery}}}\ (\bibinfo  {publisher} {SPE International Textbook, 2nd Ed.},\
  \bibinfo {year} {2018})\BibitemShut {NoStop}%
\bibitem [{\citenamefont {Paterson}(1981)}]{Paterson1981}%
  \BibitemOpen
  \bibfield  {author} {\bibinfo {author} {\bibfnamefont {L.}~\bibnamefont
  {Paterson}},\ }\bibfield  {title} {\bibinfo {title} {Radial fingering in a
  hele shaw cell},\ }\href@noop {} {\bibfield  {journal} {\bibinfo  {journal}
  {J. Fluid Mech.}\ }\textbf {\bibinfo {volume} {113}},\ \bibinfo {pages} {513}
  (\bibinfo {year} {1981})}\BibitemShut {NoStop}%
\bibitem [{\citenamefont {Paterson}(1985)}]{Paterson1985}%
  \BibitemOpen
  \bibfield  {author} {\bibinfo {author} {\bibfnamefont {L.}~\bibnamefont
  {Paterson}},\ }\bibfield  {title} {\bibinfo {title} {Fingering with miscible
  fluids in a hele shaw cell},\ }\href@noop {} {\bibfield  {journal} {\bibinfo
  {journal} {Phys. Fluids}\ }\textbf {\bibinfo {volume} {28}},\ \bibinfo
  {pages} {26} (\bibinfo {year} {1985})}\BibitemShut {NoStop}%
\bibitem [{\citenamefont {Saffman}(1986)}]{Saffman1986}%
  \BibitemOpen
  \bibfield  {author} {\bibinfo {author} {\bibfnamefont {P.~G.}\ \bibnamefont
  {Saffman}},\ }\bibfield  {title} {\bibinfo {title} {Viscous fingering in
  hele-shaw cells},\ }\href@noop {} {\bibfield  {journal} {\bibinfo  {journal}
  {J. Fluid Mech.}\ }\textbf {\bibinfo {volume} {173}},\ \bibinfo {pages} {73}
  (\bibinfo {year} {1986})}\BibitemShut {NoStop}%
\bibitem [{\citenamefont {Homsy}(1987)}]{Homsy1987}%
  \BibitemOpen
  \bibfield  {author} {\bibinfo {author} {\bibfnamefont {G.~M.}\ \bibnamefont
  {Homsy}},\ }\bibfield  {title} {\bibinfo {title} {Viscous fingering in porous
  media},\ }\href {www.annualreviews.org} {\bibfield  {journal} {\bibinfo
  {journal} {Annu. Rev. Fluid Mech.}\ }\textbf {\bibinfo {volume} {19}},\
  \bibinfo {pages} {271} (\bibinfo {year} {1987})}\BibitemShut {NoStop}%
\bibitem [{\citenamefont {Chen}(1987)}]{Chen1987}%
  \BibitemOpen
  \bibfield  {author} {\bibinfo {author} {\bibfnamefont {J.~D.}\ \bibnamefont
  {Chen}},\ }\bibfield  {title} {\bibinfo {title} {Radial viscous fingering
  patterns in hele-shaw cells},\ }\href@noop {} {\bibfield  {journal} {\bibinfo
   {journal} {Exp. Fluids}\ }\textbf {\bibinfo {volume} {5}},\ \bibinfo {pages}
  {363} (\bibinfo {year} {1987})}\BibitemShut {NoStop}%
\bibitem [{\citenamefont {Dias}\ and\ \citenamefont
  {Miranda}(2013)}]{Dias2013_2}%
  \BibitemOpen
  \bibfield  {author} {\bibinfo {author} {\bibfnamefont {E.~O.}\ \bibnamefont
  {Dias}}\ and\ \bibinfo {author} {\bibfnamefont {J.~A.}\ \bibnamefont
  {Miranda}},\ }\bibfield  {title} {\bibinfo {title} {Control of centrifugally
  driven fingering in a tapered hele-shaw cell},\ }\href@noop {} {\bibfield
  {journal} {\bibinfo  {journal} {Phys. Rev. E}\ }\textbf {\bibinfo {volume}
  {87}},\ \bibinfo {pages} {053014} (\bibinfo {year} {2013})}\BibitemShut
  {NoStop}%
\bibitem [{\citenamefont {Chevalier}\ \emph {et~al.}(2006)\citenamefont
  {Chevalier}, \citenamefont {{Ben Amar}}, \citenamefont {Bonn},\ and\
  \citenamefont {Lindner}}]{Chevalier2006}%
  \BibitemOpen
  \bibfield  {author} {\bibinfo {author} {\bibfnamefont {C.}~\bibnamefont
  {Chevalier}}, \bibinfo {author} {\bibfnamefont {M.}~\bibnamefont {{Ben
  Amar}}}, \bibinfo {author} {\bibfnamefont {D.}~\bibnamefont {Bonn}},\ and\
  \bibinfo {author} {\bibfnamefont {A.}~\bibnamefont {Lindner}},\ }\bibfield
  {title} {\bibinfo {title} {Inertial effects on saffman-taylor viscous
  fingering},\ }\href@noop {} {\bibfield  {journal} {\bibinfo  {journal} {J.
  Fluid Mech.}\ }\textbf {\bibinfo {volume} {552}},\ \bibinfo {pages} {83}
  (\bibinfo {year} {2006})}\BibitemShut {NoStop}%
\bibitem [{\citenamefont {Rocha}\ and\ \citenamefont
  {Miranda}(2013)}]{Rocha2013}%
  \BibitemOpen
  \bibfield  {author} {\bibinfo {author} {\bibfnamefont {F.~M.}\ \bibnamefont
  {Rocha}}\ and\ \bibinfo {author} {\bibfnamefont {J.~A.}\ \bibnamefont
  {Miranda}},\ }\bibfield  {title} {\bibinfo {title} {Manipulation of the
  saffman-taylor instability: A curvature-dependent surface tension approach},\
  }\href@noop {} {\bibfield  {journal} {\bibinfo  {journal} {Phys. Rev. E}\
  }\textbf {\bibinfo {volume} {87}},\ \bibinfo {pages} {013017} (\bibinfo
  {year} {2013})}\BibitemShut {NoStop}%
\bibitem [{\citenamefont {Park}\ and\ \citenamefont {Durian}(1994)}]{Park1994}%
  \BibitemOpen
  \bibfield  {author} {\bibinfo {author} {\bibfnamefont {S.~S.}\ \bibnamefont
  {Park}}\ and\ \bibinfo {author} {\bibfnamefont {D.~J.}\ \bibnamefont
  {Durian}},\ }\bibfield  {title} {\bibinfo {title} {Viscous and elastic
  fingering instabilities in foam},\ }\href@noop {} {\bibfield  {journal}
  {\bibinfo  {journal} {Phys. Rev. Lett.}\ }\textbf {\bibinfo {volume} {72}},\
  \bibinfo {pages} {3347} (\bibinfo {year} {1994})}\BibitemShut {NoStop}%
\bibitem [{\citenamefont {Coussot}(1999)}]{Coussot1999}%
  \BibitemOpen
  \bibfield  {author} {\bibinfo {author} {\bibfnamefont {P.}~\bibnamefont
  {Coussot}},\ }\bibfield  {title} {\bibinfo {title} {Saffman-taylor
  instability in yield-stress fluids},\ }\href@noop {} {\bibfield  {journal}
  {\bibinfo  {journal} {J. Fluid Mech.}\ }\textbf {\bibinfo {volume} {380}},\
  \bibinfo {pages} {363} (\bibinfo {year} {1999})}\BibitemShut {NoStop}%
\bibitem [{\citenamefont {Cardoso}\ and\ \citenamefont
  {Woods}(1995)}]{Cardoso1995}%
  \BibitemOpen
  \bibfield  {author} {\bibinfo {author} {\bibfnamefont {S.~S.}\ \bibnamefont
  {Cardoso}}\ and\ \bibinfo {author} {\bibfnamefont {A.~W.}\ \bibnamefont
  {Woods}},\ }\bibfield  {title} {\bibinfo {title} {The formation of drops
  through viscous instability},\ }\href@noop {} {\bibfield  {journal} {\bibinfo
   {journal} {J. Fluid Mech.}\ }\textbf {\bibinfo {volume} {289}},\ \bibinfo
  {pages} {351} (\bibinfo {year} {1995})}\BibitemShut {NoStop}%
\bibitem [{\citenamefont {Dias}\ \emph {et~al.}(2012)\citenamefont {Dias},
  \citenamefont {Alvarez-Lacalle}, \citenamefont {Carvalho},\ and\
  \citenamefont {Miranda}}]{Dias2012}%
  \BibitemOpen
  \bibfield  {author} {\bibinfo {author} {\bibfnamefont {E.~O.}\ \bibnamefont
  {Dias}}, \bibinfo {author} {\bibfnamefont {E.}~\bibnamefont
  {Alvarez-Lacalle}}, \bibinfo {author} {\bibfnamefont {M.~S.}\ \bibnamefont
  {Carvalho}},\ and\ \bibinfo {author} {\bibfnamefont {J.~A.}\ \bibnamefont
  {Miranda}},\ }\bibfield  {title} {\bibinfo {title} {Minimization of viscous
  fluid fingering: A variational scheme for optimal flow rates},\ }\href@noop
  {} {\bibfield  {journal} {\bibinfo  {journal} {Phys. Rev. Lett.}\ }\textbf
  {\bibinfo {volume} {109}},\ \bibinfo {pages} {144502} (\bibinfo {year}
  {2012})}\BibitemShut {NoStop}%
\bibitem [{\citenamefont {Zheng}\ \emph {et~al.}(2015)\citenamefont {Zheng},
  \citenamefont {Kim},\ and\ \citenamefont {Stone}}]{Zheng2015}%
  \BibitemOpen
  \bibfield  {author} {\bibinfo {author} {\bibfnamefont {Z.}~\bibnamefont
  {Zheng}}, \bibinfo {author} {\bibfnamefont {H.}~\bibnamefont {Kim}},\ and\
  \bibinfo {author} {\bibfnamefont {H.~A.}\ \bibnamefont {Stone}},\ }\bibfield
  {title} {\bibinfo {title} {Controlling viscous fingering using time-dependent
  strategies},\ }\href {https://doi.org/10.1103/PhysRevLett.115.174501}
  {\bibfield  {journal} {\bibinfo  {journal} {Phys. Rev. Lett.}\ }\textbf
  {\bibinfo {volume} {115}},\ \bibinfo {pages} {174501} (\bibinfo {year}
  {2015})}\BibitemShut {NoStop}%
\bibitem [{\citenamefont {Pihler-Puzović}\ \emph {et~al.}(2012)\citenamefont
  {Pihler-Puzović}, \citenamefont {Illien}, \citenamefont {Heil},\ and\
  \citenamefont {Juel}}]{Pihler2012}%
  \BibitemOpen
  \bibfield  {author} {\bibinfo {author} {\bibfnamefont {D.}~\bibnamefont
  {Pihler-Puzović}}, \bibinfo {author} {\bibfnamefont {P.}~\bibnamefont
  {Illien}}, \bibinfo {author} {\bibfnamefont {M.}~\bibnamefont {Heil}},\ and\
  \bibinfo {author} {\bibfnamefont {A.}~\bibnamefont {Juel}},\ }\bibfield
  {title} {\bibinfo {title} {Suppression of complex finger-like patterns at the
  interface between air and a viscous fluid by elastic membranes},\ }\href@noop
  {} {\bibfield  {journal} {\bibinfo  {journal} {Phys. Rev. Lett.}\ }\textbf
  {\bibinfo {volume} {108}},\ \bibinfo {pages} {074502} (\bibinfo {year}
  {2012})}\BibitemShut {NoStop}%
\bibitem [{\citenamefont {Pihler-Puzović}\ \emph {et~al.}(2013)\citenamefont
  {Pihler-Puzović}, \citenamefont {Périllat}, \citenamefont {Russell},
  \citenamefont {Juel},\ and\ \citenamefont {Heil}}]{Pihler2013}%
  \BibitemOpen
  \bibfield  {author} {\bibinfo {author} {\bibfnamefont {D.}~\bibnamefont
  {Pihler-Puzović}}, \bibinfo {author} {\bibfnamefont {R.}~\bibnamefont
  {Périllat}}, \bibinfo {author} {\bibfnamefont {M.}~\bibnamefont {Russell}},
  \bibinfo {author} {\bibfnamefont {A.}~\bibnamefont {Juel}},\ and\ \bibinfo
  {author} {\bibfnamefont {M.}~\bibnamefont {Heil}},\ }\bibfield  {title}
  {\bibinfo {title} {Modelling the suppression of viscous fingering in
  elastic-walled hele-shaw cells},\ }\href@noop {} {\bibfield  {journal}
  {\bibinfo  {journal} {J. Fluid Mech.}\ }\textbf {\bibinfo {volume} {731}},\
  \bibinfo {pages} {162} (\bibinfo {year} {2013})}\BibitemShut {NoStop}%
\bibitem [{\citenamefont {Al-Housseiny}\ \emph {et~al.}(2013)\citenamefont
  {Al-Housseiny}, \citenamefont {Christov},\ and\ \citenamefont
  {Stone}}]{Housseiny2013}%
  \BibitemOpen
  \bibfield  {author} {\bibinfo {author} {\bibfnamefont {T.~T.}\ \bibnamefont
  {Al-Housseiny}}, \bibinfo {author} {\bibfnamefont {I.~C.}\ \bibnamefont
  {Christov}},\ and\ \bibinfo {author} {\bibfnamefont {H.~A.}\ \bibnamefont
  {Stone}},\ }\bibfield  {title} {\bibinfo {title} {Two-phase fluid
  displacement and interfacial instabilities under elastic membranes},\
  }\href@noop {} {\bibfield  {journal} {\bibinfo  {journal} {Phys. Rev. Lett.}\
  }\textbf {\bibinfo {volume} {111}},\ \bibinfo {pages} {034502} (\bibinfo
  {year} {2013})}\BibitemShut {NoStop}%
\bibitem [{\citenamefont {Al-Housseiny}\ \emph {et~al.}(2012)\citenamefont
  {Al-Housseiny}, \citenamefont {Tsai},\ and\ \citenamefont
  {Stone}}]{Housseiny2012}%
  \BibitemOpen
  \bibfield  {author} {\bibinfo {author} {\bibfnamefont {T.~T.}\ \bibnamefont
  {Al-Housseiny}}, \bibinfo {author} {\bibfnamefont {P.~A.}\ \bibnamefont
  {Tsai}},\ and\ \bibinfo {author} {\bibfnamefont {H.~A.}\ \bibnamefont
  {Stone}},\ }\bibfield  {title} {\bibinfo {title} {Control of interfacial
  instabilities using flow geometry},\ }\href@noop {} {\bibfield  {journal}
  {\bibinfo  {journal} {Nat. Phys.}\ }\textbf {\bibinfo {volume} {8}},\
  \bibinfo {pages} {747} (\bibinfo {year} {2012})}\BibitemShut {NoStop}%
\bibitem [{\citenamefont {Al-Housseiny}\ and\ \citenamefont
  {Stone}(2013)}]{Stone2013}%
  \BibitemOpen
  \bibfield  {author} {\bibinfo {author} {\bibfnamefont {T.~T.}\ \bibnamefont
  {Al-Housseiny}}\ and\ \bibinfo {author} {\bibfnamefont {H.~A.}\ \bibnamefont
  {Stone}},\ }\bibfield  {title} {\bibinfo {title} {Controlling viscous
  fingering in tapered hele-shaw cells},\ }\href@noop {} {\bibfield  {journal}
  {\bibinfo  {journal} {Phys. Fluids}\ }\textbf {\bibinfo {volume} {25}},\
  \bibinfo {pages} {092102} (\bibinfo {year} {2013})}\BibitemShut {NoStop}%
\bibitem [{\citenamefont {Bongrand}\ and\ \citenamefont
  {Tsai}(2018)}]{Bongrand2018}%
  \BibitemOpen
  \bibfield  {author} {\bibinfo {author} {\bibfnamefont {G.}~\bibnamefont
  {Bongrand}}\ and\ \bibinfo {author} {\bibfnamefont {P.~A.}\ \bibnamefont
  {Tsai}},\ }\bibfield  {title} {\bibinfo {title} {Manipulation of viscous
  fingering in a radially tapered cell geometry},\ }\href@noop {} {\bibfield
  {journal} {\bibinfo  {journal} {Phys. Rev. E}\ }\textbf {\bibinfo {volume}
  {97}},\ \bibinfo {pages} {061101} (\bibinfo {year} {2018})}\BibitemShut
  {NoStop}%
\bibitem [{\citenamefont {Mirzadeh}\ and\ \citenamefont
  {Bazant}(2017)}]{Mirzadeh2017}%
  \BibitemOpen
  \bibfield  {author} {\bibinfo {author} {\bibfnamefont {M.}~\bibnamefont
  {Mirzadeh}}\ and\ \bibinfo {author} {\bibfnamefont {M.~Z.}\ \bibnamefont
  {Bazant}},\ }\bibfield  {title} {\bibinfo {title} {Electrokinetic control of
  viscous fingering},\ }\href@noop {} {\bibfield  {journal} {\bibinfo
  {journal} {Phys. Rev. Lett.}\ }\textbf {\bibinfo {volume} {119}},\ \bibinfo
  {pages} {174501} (\bibinfo {year} {2017})}\BibitemShut {NoStop}%
\bibitem [{\citenamefont {Herschel}\ and\ \citenamefont
  {Bulkley}(1926)}]{Herschel1926}%
  \BibitemOpen
  \bibfield  {author} {\bibinfo {author} {\bibfnamefont {W.~H.}\ \bibnamefont
  {Herschel}}\ and\ \bibinfo {author} {\bibfnamefont {R.}~\bibnamefont
  {Bulkley}},\ }\bibfield  {title} {\bibinfo {title} {Konsistenzmessungen von
  gummi-benzollösungen},\ }\href@noop {} {\bibfield  {journal} {\bibinfo
  {journal} {Kolloid Z.}\ }\textbf {\bibinfo {volume} {39}},\ \bibinfo {pages}
  {291} (\bibinfo {year} {1926})}\BibitemShut {NoStop}%
\bibitem [{\citenamefont {Maleki-Jirsaraei}\ \emph {et~al.}(2005)\citenamefont
  {Maleki-Jirsaraei}, \citenamefont {Lindner}, \citenamefont {Rouhani},\ and\
  \citenamefont {Bonn}}]{Jirsaraei2005}%
  \BibitemOpen
  \bibfield  {author} {\bibinfo {author} {\bibfnamefont {N.}~\bibnamefont
  {Maleki-Jirsaraei}}, \bibinfo {author} {\bibfnamefont {A.}~\bibnamefont
  {Lindner}}, \bibinfo {author} {\bibfnamefont {S.}~\bibnamefont {Rouhani}},\
  and\ \bibinfo {author} {\bibfnamefont {D.}~\bibnamefont {Bonn}},\ }\bibfield
  {title} {\bibinfo {title} {Saffman–taylor instability in yield stress
  fluids},\ }\href@noop {} {\bibfield  {journal} {\bibinfo  {journal} {J.
  Phys.: Condens. Matter}\ }\textbf {\bibinfo {volume} {17}},\ \bibinfo {pages}
  {1219} (\bibinfo {year} {2005})}\BibitemShut {NoStop}%
\bibitem [{\citenamefont {Eslami}\ and\ \citenamefont
  {Taghavi}(2017)}]{Eslami2017}%
  \BibitemOpen
  \bibfield  {author} {\bibinfo {author} {\bibfnamefont {A.}~\bibnamefont
  {Eslami}}\ and\ \bibinfo {author} {\bibfnamefont {S.~M.}\ \bibnamefont
  {Taghavi}},\ }\bibfield  {title} {\bibinfo {title} {Viscous fingering regimes
  in elasto-visco-plastic fluids},\ }\href@noop {} {\bibfield  {journal}
  {\bibinfo  {journal} {J. Non-Newt. Fluid Mech.}\ }\textbf {\bibinfo {volume}
  {243}},\ \bibinfo {pages} {79} (\bibinfo {year} {2017})}\BibitemShut
  {NoStop}%
\bibitem [{\citenamefont {Eslami}\ and\ \citenamefont
  {Taghavi}(2019)}]{Eslami2019}%
  \BibitemOpen
  \bibfield  {author} {\bibinfo {author} {\bibfnamefont {A.}~\bibnamefont
  {Eslami}}\ and\ \bibinfo {author} {\bibfnamefont {S.~M.}\ \bibnamefont
  {Taghavi}},\ }\bibfield  {title} {\bibinfo {title} {Viscous fingering of
  yield stress fluids: The effects of wettability},\ }\href@noop {} {\bibfield
  {journal} {\bibinfo  {journal} {J. Non-Newt. Fluid Mech.}\ }\textbf {\bibinfo
  {volume} {264}},\ \bibinfo {pages} {25} (\bibinfo {year} {2019})}\BibitemShut
  {NoStop}%
\bibitem [{\citenamefont {Eslami}\ \emph {et~al.}(2020)\citenamefont {Eslami},
  \citenamefont {Basak},\ and\ \citenamefont {Taghavi}}]{Eslami2020_1}%
  \BibitemOpen
  \bibfield  {author} {\bibinfo {author} {\bibfnamefont {A.}~\bibnamefont
  {Eslami}}, \bibinfo {author} {\bibfnamefont {R.}~\bibnamefont {Basak}},\ and\
  \bibinfo {author} {\bibfnamefont {S.~M.}\ \bibnamefont {Taghavi}},\
  }\bibfield  {title} {\bibinfo {title} {Multiphase viscoplastic flows in a
  nonuniform hele-shaw cell: A fluidic device to control interfacial
  patterns},\ }\href@noop {} {\bibfield  {journal} {\bibinfo  {journal} {Ind.
  Eng. Chem. Res.}\ }\textbf {\bibinfo {volume} {59}},\ \bibinfo {pages} {4119}
  (\bibinfo {year} {2020})}\BibitemShut {NoStop}%
\end{thebibliography}%

\appendix

\section*{Methods}
{\bf Sample Preparation}~~~~The two aqueous solutions of PAA (SigmaAldrich, $M_w \approx 1,250,000$) are prepared to produce different viscosity contrasts. Both solutions have the same polymer concentration, by slowly adding the polymer powder in water and subsequently stirring the mixture at high speed for 1~hr. The mixture generates an acid solution that can be neutralized using a basic solution. The two PAA solutions are prepared with (S1) and without \ce{NaOH} (S2), stirred for another 10 hours at medium speed. Finally, after the agitation, the solution is allowed to rest for a day before performing rheological measurements. 

\begin{figure}[h!]
    \begin{center}
    \includegraphics[width=3.2in]{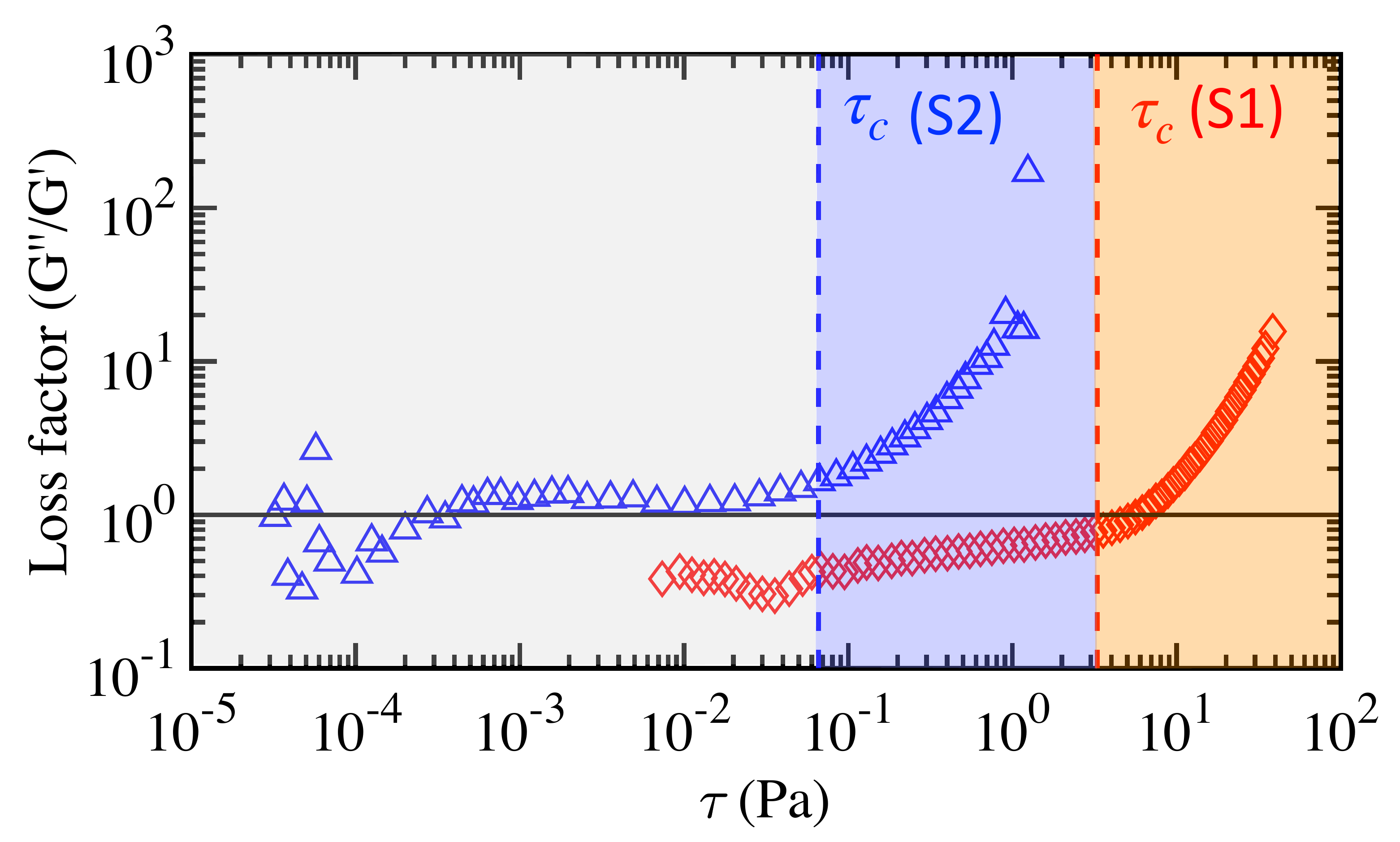}
    \end{center}
    \protect{\caption{The data of loss factor, the ratio of the loss modulus ($G''$) to the storage modulus ($G'$), varying with the shear stress, $\tau$, obtained during oscillation amplitude sweep test at constant frequency ($\hat{w} = 1$ rad/s). The vertical dashed lines represent the yield-stress ($\tau_c$) values of the two fluids.}
    \label{Fig4}}
\end{figure}

{\bf Rheological Measurements}~~~~We further perform oscillation amplitude sweep tests at constant frequency ($\hat{\omega}$ = 1~rad/s) to validate negligible elasticity of the complex fluids. Shown in Fig.~\ref{Fig4} below are the results of the loss factor, i.e., the ratio of the loss modulus ($G''$) to the storage modulus ($G'$). The former $G''$ represents the viscous properties of the complex fluids, while the latter $G'$ fluid elasticity with respect to the shear stress, $\tau$. The fluid's viscous behavior prevails when the loss factor ($G''/G'$) is greater than unity, whereas elastic for $G''/G' < 1$.  The vertical dashed lines represents the yield-stress values for the fluids (S1) and (S2). We only focus on the (color shaded) regime whereby fluids are flowing, i.e, $\tau > \tau_c$, when $G''/G' \gtrsim 1$, meaning the viscous component prevails.  Hence we are able to neglect the elastic effects of the fluids in the theoretical analysis.

\end{document}